\documentclass[a4paper,journal]{IEEEtran}
\usepackage{amsmath,amsfonts}
\usepackage{algorithmic}
\usepackage{array}
\usepackage[caption=false,font=normalsize,labelfont=sf,textfont=sf]{subfig}
\usepackage{xcolor, graphicx}
\usepackage{cite}
\usepackage{url}
\usepackage{color}
\usepackage{hyperref} 
\usepackage{blindtext}

\usepackage{amssymb}% http://ctan.org/pkg/amssymb
\usepackage{pifont}% http://ctan.org/pkg/pifont
\usepackage{dirtytalk}

\begin{document}

\title{Offline Multi-Agent Reinforcement Learning for 6G Communications: Fundamentals,
Applications and Future Directions
} 

\author{Eslam Eldeeb and Hirley Alves
	\thanks{This work was supported by 6G Flagship (Grant Number 369116) funded by the Research Council of Finland.}
	
	\thanks{The authors are with Centre for Wireless Communications (CWC), University of Oulu, Finland. Email: firstname.lastname@oulu.fi. 
}
}

\maketitle
\maketitle

\begin{abstract}
The next-generation wireless technologies, including beyond 5G and 6G networks, are paving the way for transformative applications such as vehicle platooning, smart cities, and remote surgery. These innovations are driven by a vast array of interconnected wireless entities, including IoT devices, access points, UAVs, and CAVs, which increase network complexity and demand more advanced decision-making algorithms. Artificial intelligence (AI) and machine learning (ML), especially reinforcement learning (RL), are key enablers for such networks, providing solutions to high-dimensional and complex challenges. However, as networks expand to multi-agent environments, traditional online RL approaches face cost, safety, and scalability limitations. Offline multi-agent reinforcement learning (MARL) offers a promising solution by utilizing pre-collected data, reducing the need for real-time interaction. This article introduces a novel offline MARL algorithm based on conservative Q-learning (CQL), ensuring safe and efficient training. We extend this with meta-learning to address dynamic environments and validate the approach through use cases in radio resource management and UAV networks. Our work highlights offline MARL's advantages, limitations, and future directions in wireless applications.
\end{abstract}

%\MS{current number of words:700+1800+700+400+100=3700}

\section{Introduction} \label{sec:intro} 
%\MS{700 words}

The next generation of wireless technology and the 6G era bring about advanced use cases such as vehicle platooning, smart cities, and remote surgeries~\cite{9144301}. These applications rely on the extensive deployment of diverse wireless entities, including low-power IoT devices, base stations (BSs), unmanned aerial vehicles (UAVs), and connected and autonomous vehicles (CAVs). This progress is marked by increased network complexity and a rise in non-linear parameters. Consequently, the next generation of wireless technology necessitates efficient decision-making algorithms to replace traditional methods, which often fall short of meeting the demands of these emerging applications.

Artificial intelligence (AI) and machine learning (ML) are poised to become key enablers of the next-generation 5G-beyond and 6G networks. Leveraging advanced deep neural networks and powerful function approximators, these technologies can tackle complex, high-dimensional challenges. Reinforcement learning (RL) stands out as a decision-making powerhouse within the ML domain. Its adaptability makes it an ideal fit for wireless networks, as it thrives in sequential environments broken down into discrete time steps~\cite{9738819}. Notably, RL employs model-free algorithms, enabling it to navigate and solve intricate network scenarios without relying on rigid, closed-form models~\cite{10155733}.

\begin{figure*}[t!]
    \centering    \includegraphics[width=1.65\columnwidth,trim={0cm 0 0cm 0},clip]{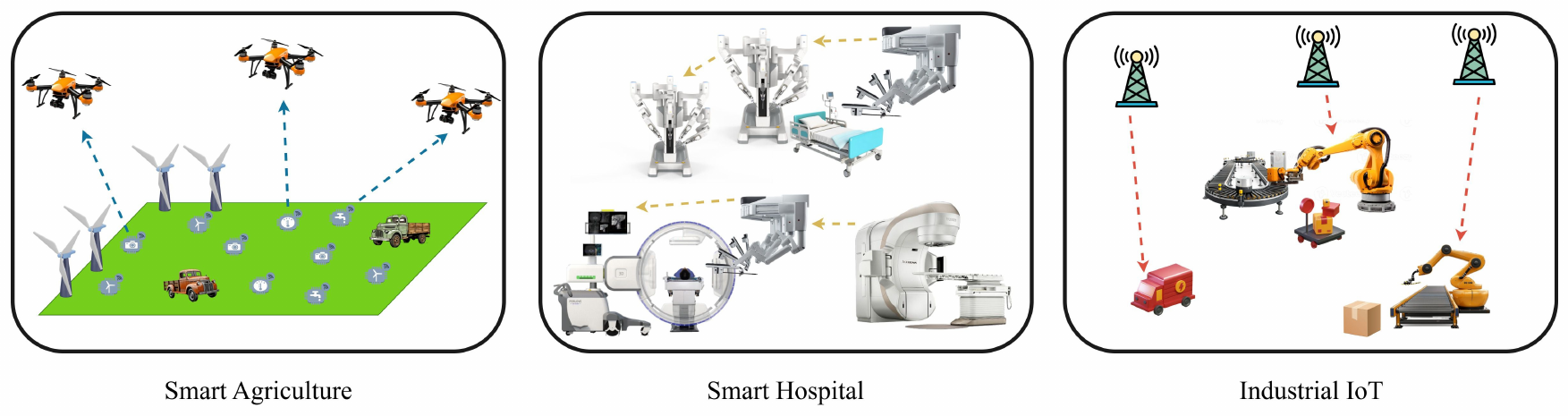} \vspace{2mm}
    \caption{Illustration of emerging applications in $6$G network with multiple agents. This includes smart agriculture, smart hospital, and industrial IoT.}
    \vspace{0mm}
    \label{Applications}
\end{figure*}

As wireless networks continue to expand and support an ever-growing array of applications, RL has evolved from single-agent systems to multi-agent RL (MARL), where multiple entities interact within the same environment~\cite{9372298,8807386}. For instance, imagine a swarm of UAVs supporting a vast deployment of IoT devices in a smart agriculture setting, as depicted in Fig.~\ref{Applications}. Similarly, smart hospitals are expected to incorporate autonomous robots for surgeries and medical procedures, transforming healthcare delivery. Another promising use case lies in industrial IoT, where networks of autonomous robots and sensors linked via multiple access points collaborate on complex tasks and measurements. To meet the demands of these scenarios, efficient radio resource management (RRM) is crucial.

While MARL algorithms are well-suited for tackling these advanced applications, their real-world deployment remains confined to research settings. Most MARL algorithms developed for the wireless domain, including off-policy deep MARL methods, operate as online algorithms that depend heavily on continuous environmental interaction. However, this approach presents significant challenges~\cite{10753476}. Online data collection is both time-consuming and costly, especially in multi-agent scenarios where the volume of required data scales dramatically. Additionally, online interactions can be inherently unsafe, as agents might make exploratory decisions that could result in critical failures or disastrous outcomes.

Offline MARL has emerged as a promising alternative to traditional online MARL. This approach optimizes policies using a static offline dataset, eliminating the need for real-time interactions with the environment. Such datasets are typically generated through behavioral policies. However, a key challenge with offline MARL is the distributional shift, the mismatch between the experiences in the dataset and those learned by the algorithm. Fortunately, offline MARL algorithms, such as conservative Q-learning (CQL), address this issue by minimizing the influence of unseen experiences during the learning process~\cite{levine2020offl}. In multi-agent scenarios, offline MARL proves invaluable, significantly reducing time and computational demands by leveraging pre-collected data.

In this article, we investigate the potential of offline MARL for next-generation wireless networks, focusing on its ability to ensure safe training without costly online interactions. We extend CQL to the multi-agent setting, addressing distributional shift, coordination, and computational trade-offs that are critical in wireless environments. Furthermore, we integrate meta-learning (MAML) with offline MARL to enable rapid adaptation in dynamic scenarios where network objectives or conditions change. The proposed framework is validated through case studies on radio resource management and UAV trajectory planning, demonstrating its practicality and effectiveness. The main contributions of this work are as follows:
\begin{itemize}
    \item We present a detailed overview of RL and MARL in wireless networks, highlighting their limitations for real-world deployment, particularly in terms of safety, scalability, and data collection.

    \item We extend CQL to multi-agent scenarios through both independent and CTDE formulations, addressing distributional shift, coordination, and computational trade-offs specific to wireless systems.
    
    \item We integrate meta-learning via MAML with offline MARL to enhance scalability and generalization across dynamic environments, enabling faster adaptation under varying objectives.
    
    \item We validate the proposed framework through wireless-specific case studies in radio resource management and UAV trajectory planning, and conclude with open challenges and future research directions.
\end{itemize}

\section{Multi-Agent Reinforcement Learning} \label{sec:RL}
%\MS{1800 words}
\subsection{RL}
In conventional RL frameworks, problems are formulated as Markov decision processes (MDPs), where an agent surfs the environment, observes the state of the environment, takes an action, and then receives a reward accordingly from the environment that evaluates the action taken at the observed state. The agent gets a high reward for good selected actions, whereas it receives a penalty if bad actions are chosen. A policy is a strategy that defines the agent behavior in the environment, \emph{i.e.}, what action to select at each state. The agent's objective is to find the optimum policy that maximizes the return (cumulative rewards) over time. The state-action value function (known as Q-function) is a function (look-up table) that estimates the return of taking a specific action at a state and following the policy onward.

To find the optimal policy, the agent needs to solve the optimal Q-function that estimates the actual return of the state-action pairs. To this end, Q-learning, a model-free RL algorithm, solves the optimal Q-function iteratively using Bellman theories. Q-learning relies on temporal difference (TD) by visiting and updating the Q-function each time step with a practical learning rate. Although it's highly applicable in solving various MDPs, it stands short when the dimension of the environment increases due to its need to visit most of the state-action pairs in the environment.

Deep RL is an emerging RL variant that combines traditional RL algorithms, such as Q-learning, with deep learning, such as deep neural networks, to overcome the dimensionality curse of large MDPs. Deep RL uses deep neural networks as function approximators to model and estimate the Q-function. Deep Q-networks (DQNs) revolutionized the RL field due to their efficiency in solving large-dimension games like Go and Atari. They rely on off-policy techniques by saving the visited experiences in a replay memory and sampling them in future time steps to contribute to the true Q-function estimation. DQNs have been a primary element in solving most recent wireless communication problems, such as UAV networks, RRM, and federated RL-based approaches in cloud–edge collaborative IoT~\cite{10173668}.

\subsection{MARL}
With the progress of wireless communication entities and the emergence of new and complex applications, wireless networks urge the need for distributed and multi-agent solutions. In the multi-agent setting, RL problems evolve into MARL problems, which are typically formulated as multi-agent-MDP (MA-MDP). In MA-MDP, multiple agents surf the environment, observe their observations and subsets of the whole environment state, take action, and receive a reward that evaluates the action taken \cite{9084325}.

\begin{figure*}[t!]
    \centering    \includegraphics[width=1.65\columnwidth,trim={0cm 0 0cm 0},clip]{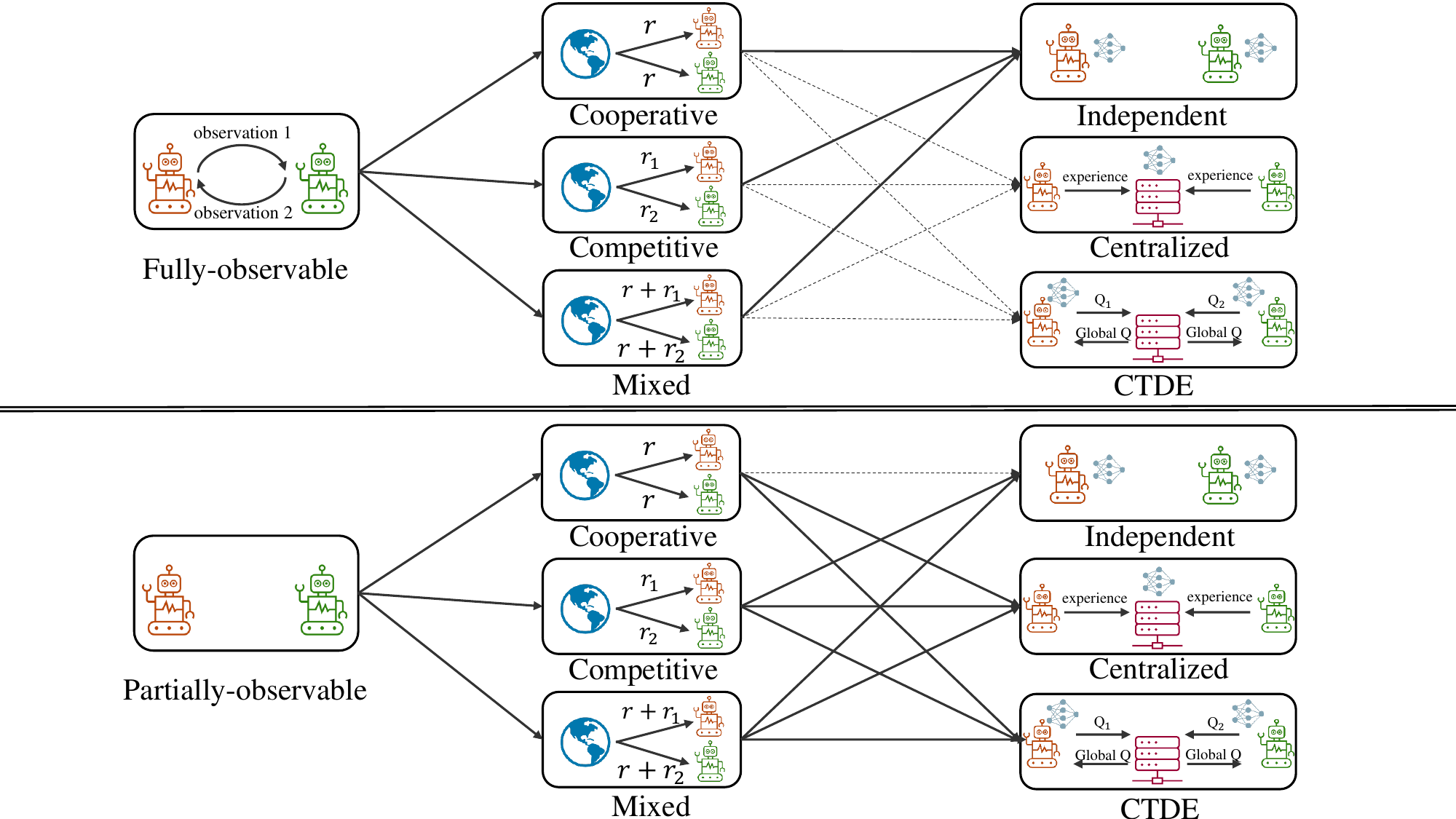} \vspace{2mm}
    \caption{Illustration of different MARL frameworks. The MARL problem varies according to the environment state, objective, and algorithm execution. Solid lines indicate the popularity of combining MARL schemes, while dotted lines indicate the popularity of the less popular ones. This figure also serves as a system-level conceptual illustration of how MARL training can be structured in wireless and 6G environments.}
    \vspace{0mm}
    \label{MARL_Variants}
\end{figure*}

The MARL problem varies according to three parameters:
\begin{enumerate}
    \item \textit{Environment state:} The MARL problem is classified according to the observations seen by each agent:
    \begin{itemize}
        \item \textbf{Fully-observable MARL:} Each agent has full access to the overall state of the environment, \emph{i.e.}, each agent contemplates its observation and the observations of other agents.

        \item \textbf{Partially-observable  MARL:} Each agent observes its own internal observations only; and
    \end{itemize}

    \item \textit{Objective:} The MARL problem is defined according to the overall objective of the agents in the environment:
    \begin{itemize}
        \item \textbf{Cooperative MARL:} All agents cooperate towards one unified objective. Typically, all agents receive the same reward, which is calculated in a central unit based on the actions of all agents.

        \item \textbf{Competitive MARL:} Agents compete to maximize their gain. Each agent receives individual rewards depending on their actions.

        \item \textbf{Mixed cooperative and competitive MARL:} Agents cooperate towards a common objective while aiming for their gains. Rewards comprise \textit{i)} the common reward part and \textit{ii)} the individual gain;
    \end{itemize}
    
    \item \textit{Algorithm execution:} According to this criteria, the MARL problem can be classified into three groups:
    \begin{itemize}
        \item \textbf{Independent MARL:} Each agent optimizes its policy / Q-function individually. The learning is performed internally at each agent.

        \item \textbf{Centralized training decentralized execution (CTDE) MARL:} During training, each agent shares his Q-function with a central unit that estimates the global Q-function using the individual Q-functions. During testing, each agent follows its Q-function internally.
        
        \item \textbf{Centralized MARL:} All states and actions are concatenated, and one global Q-function is optimized, comprising the individual Q-functions. The learning is performed at a central unit.
    \end{itemize}
\end{enumerate}
It is worth mentioning that the CTDE framework is commonly used with partially observable MARL and cooperative MARL to overcome the limited information of the agents about other agents' observations. Meanwhile, Independent MARL is commonly used with fully-observable and cooperative MARL or partially-observable MARL and competitive MARL. Fig.~\ref{MARL_Variants} summarizes MARL variants. Like the single-agent DQN, each Q-network is modeled as a deep neural network in the multi-agent DQN (MA-DQN). Beyond taxonomy, Fig.~\ref{MARL_Variants} provides a system-level conceptual view of how MARL training modes can be structured in wireless networks, serving as a high-level guide for 6G deployment scenarios.

To this end, fully-observable MARL is more efficient than partially-observable MARL, but at the cost of higher signaling overhead due to the state information sharing. Similarly, centralized MARL is more efficient than independent MARL due to the global Q-function trained in the centralized case. However, independent MARL consumes lower communication and computational resources due to the low signaling overhead and the training division among agents. CTDE MARL is more efficient than independent MARL and consumes fewer resources than centralized MARL. Finally, cooperative MARL is more common in wireless networks due to the common objective in the wireless environments compared to the competitive and mixed cooperative and competitive MARL, which are usually more common in games.

%\begin{table}[!t]
%\centering
%    \caption{A comparison between online RL, off-policy deep RL, offline RL, distributional RL and offline distributional RL.
%}
%\label{MARL_Comparison}
%\begin{tabular}{|c|c|c|c|}
%\hline
% & \textbf{State} & \textbf{Reward} & \textbf{Q-function}\\ 
%\hline
%\textbf{Fully-observable} & full state & - & -\\
%\hline

%\textbf{Partially-observable} & internal observations & - & -\\
%\hline
%\hline

%\textbf{Cooperative} & - & same & -\\
%\hline                              

%\textbf{Competitive} & - & different & -\\
%\hline                              

%\textbf{Mixed} & - & mixed & -\\
%\hline
%\hline

%\textbf{Independent} & - & - & multiple\\
%\hline                              

%\textbf{CTDE} & - & - & multiple\\
%\hline                              

%\textbf{Centralized} & - & - & single\\
%\hline
%\end{tabular} \vspace{-0mm}
%\end{table}

Despite the significant breakthrough in developing efficient MARL algorithms for each MARL variant, it falls short in real-time wireless communication applications. First, real-time wireless environments are often stochastic and full of uncertainties. Hence, they require expensive data collection by interacting with the environment online, which might be unsafe or unfeasible. Second, real-time wireless environments are dynamic, where network configurations change, such as channel characteristics, number of devices, and network objectives. These challenges urge new MARL algorithms that can be adopted safely and efficiently in the real world.

\section{Offline MARL} \label{sec:ODRL}
%\MS{700 words}
\subsection{Offline MARL}
Offline MARL is a subset of MARL that learns the optimal policies using an offline, fixed dataset without any online interaction with the environment. The offline dataset is usually collected using behavioral policies, such as any online MARL algorithm, classic algorithms, or random policies. Deploying MARL algorithms like MA-DQN using offline datasets generally fails. This failure occurs due to the distributional shift between the policies collected in the dataset and the learned policy from the Q-network estimation. This shift creates out-of-distribution (OOD) actions, which are actions that are not sufficiently covered in the offline dataset, introducing optimistic overestimation~\cite{NEURIPS2022_01d78b29}. In online MA-DQN, the OOD actions collected in the replay buffers are corrected by online interaction with the environment and truly estimating the Q-values for these actions. This is not the case for offline MA-DQN due to the lack of online interactions with the environment.

Recent advances in offline RL and offline MARL address the distributional shift problem by explicitly constraining actions as in-distribution. In other words, they penalize the OOD actions to limit their weight on the Q-function update. CQL is an offline RL/MARL algorithm that adds a regularization parameter to the reward so that OOD actions sustain a more significant loss than in-distribution actions~\cite{kumar2020conservative}. CQL is easy to implement on top of DQNs for any MARL variant. In addition to CQL, simpler methods such as behavior cloning (BC) directly learn policies by imitating the actions in the dataset, offering stability at the cost of limited exploration.  While not the main focus of this work, these methods complement CQL-based approaches and broaden the toolbox of safe offline MARL.

Compared to online MARL, offline MARL significantly reduces sample complexity by avoiding costly real-time exploration and scales more effectively as the number of agents grows. However, offline MARL remains dependent on the diversity and quality of pre-collected datasets, whereas online MARL benefits from continuous hardware feedback to adapt in real time. This trade-off highlights the complementary strengths of the two paradigms.

\subsection{Meta-Learning}

Meta-learning is a branch of machine learning, known as learning to learn, which aims to enhance learning ability in new, unseen environments. Meta-learning relies on learning across different tasks, which are assumed to be drawn from a single task distribution, to minimize the data and computation needed to learn a new, unseen task drawn from the same distribution. Hence, generalizing learning across multiple tasks rather than task-specific optimization enables quick adaptation with a few iterations and a small amount of data. Therefore, meta-learning minimizes the needed training intervals, required computational power, and data size.

MAML is a well-known meta-learning algorithm that optimizes the initial parameters of a deep neural network, such as a Q-network, enabling fast adaptation to optimal parameters across different tasks or environments~\cite{FINN}. Multiple tasks are sampled during meta-training, and the initial parameters are updated through task-specific adaptation and meta-optimization loops. A new unseen task is sampled during meta-testing, and the converged initial parameters are updated over a few iterations. MAML enables few-shot learning, where there are few shots of data in the meta-testing phase, and zero-shot learning, where no data is available.

In the multi-agent setting, integrating offline MAML poses additional theoretical and practical considerations compared to the single-agent case. First, the inherent non-stationarity of MARL means that each agent's learning depends on others' evolving policies, further amplifying the challenge of task generalization. By initializing the Q-network parameters to optimize across multiple related tasks, MAML mitigates this non-stationarity and accelerates convergence to effective policies even under shifting network conditions. Second, the task distribution in dynamic wireless environments, such as varying channel conditions or mobility patterns, may deviate significantly. In such cases, the initialization still offers a favorable bias that supports few-shot adaptation and better robustness against distributional shifts. This highlights MAML's ability to encode transferable knowledge across dynamic MARL tasks.

In summary, our framework addresses several limitations of prior approaches. Unlike online MARL methods that rely on continuous, often unsafe exploration, the proposed offline MARL method relies solely on pre-collected datasets, ensuring safer training. Compared to classical offline RL methods, we extend CQL to multi-agent environments, mitigating distributional shift and coordination challenges. Finally, by integrating meta-learning, our framework adapts quickly to dynamic environments, a property not directly supported by existing MARL or offline RL approaches. Particularly, the proposed meta-training process explicitly learns the initialization parameters by training across multiple related tasks, where each task-specific policy contributes to refining a shared parameter set that generalizes across environments. This learned initialization enables rapid convergence and robust adaptation when agents face new, unseen objectives or network conditions. Together, these features position our method as a safe, scalable, and adaptable solution for 6G networks.

\section{Selected Applications}\label{sec:Applications}
In this section, we evaluate the proposed offline MARL algorithm based on the CQL framework across three distinct use cases: \textit{i)} radio resource management, \textit{ii)} UAV trajectory planning, and \textit{iii)} UAV trajectory planning in dynamic networks. Each use case represents a partially observable, cooperative MARL scenario in which agents collaborate towards a shared objective, relying solely on their local observations. The Q-functions in the proposed CQL algorithms and in all use cases are implemented as neural networks comprising two hidden layers, each with $256$ neurons. We benchmark the performance of two variations of the proposed offline MARL approach, independent CQL (I-CQL) and centralized training with decentralized execution CQL (CTDE-CQL), against baseline methods, including classical methods and offline MARL methods, such as DQN and batch-constrained Q-learning (BCQ). To enhance the timeliness of the collected datasets in all experiments, each episode includes randomized device locations, mobility patterns, and channel conditions, providing diversity that reflects non-stationary environments.

\begin{figure}[t!]
    \centering    \includegraphics[width=0.8\columnwidth,trim={0cm 0 0cm 0},clip]{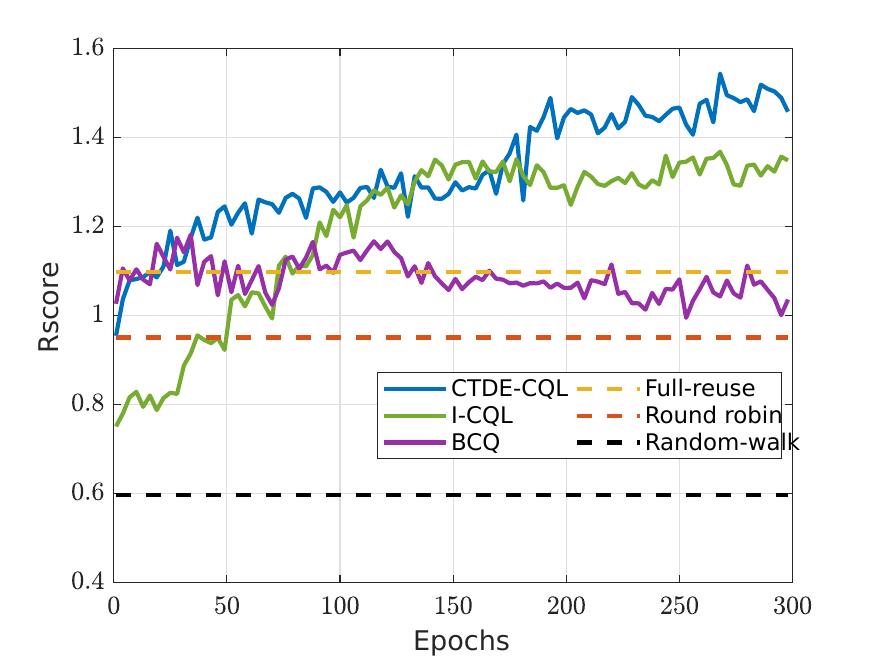} \vspace{2mm}
    \caption{Average Rscore over $100$ unique test episodes as a function of the number of training epochs.}
    \vspace{0mm}
    \label{RRM_Fig}
\end{figure}

\subsection{Case Study 1: Radio Resource Management} \label{sec:RRM_usecase}
We simulate a $5000 \: \text{m} \times 5000 \: \text{m}$ square area containing $4$ operating access points (APs) and $24$ user equipments (UEs) over $2000$ discrete time steps. In all episodes, UE / AP positions are randomized, which reflects mobility and heterogeneity in practical deployments. Each UE, moving randomly at a constant speed of $1 \: \text{m/s}$, is assigned to one AP at the start of each episode. The APs rank their associated UEs based on the proportional fairness (PF) factor, which reflects the quality of service and data rates experienced by the UEs. Each AP observes the signal-to-interference-plus-noise ratio (SINR) and PF values for its top $3$ associated UEs (local observations) and selects one UE from this group to serve (action). The APs collaborate to maximize the $R$-score, defined as a weighted sum of the total sum rate and the $5$th-percentile rate of the UEs. Consequently, a universal reward is formulated as the sum of the weighted PF and instantaneous rates across all UEs~\cite{eldeeb2025offlinemultiagentreinforcementlearning}. We define the R-score as a weighted combination of two terms: the total sum rate across all users and the 5th-percentile user rate, which captures fairness among users. This metric balances throughput and fairness in evaluating RRM performance.

We generate an offline dataset using $20 \%$ of the experience obtained from an online soft actor-critic (SAC) algorithm. That ratio was chosen to emulate realistic offline data collection, where gathering large-scale online interactions is costly and potentially unsafe. This choice also illustrates that the proposed method can perform well under diverse data quality and limited data availability. We evaluate the proposed approach against three well-known benchmarks: full-reuse, which always selects the device with the highest PF factor; round robin, which allocates resources evenly among users; and random-walk, which selects actions randomly. As illustrated in Fig.~\ref{RRM_Fig}, the CTDE-CQL algorithm outperforms all baseline methods, including I-CQL and BCQ. This is attributed to CTDE-CQL's reliance on Q-function sharing between agents, which improves the policy optimization of individual agents. In contrast, I-CQL does not leverage the experiences of other agents, leading to degraded training performance.

\begin{figure}[t!]
    \centering    \includegraphics[width=0.8\columnwidth,trim={0cm 0 0cm 0},clip]{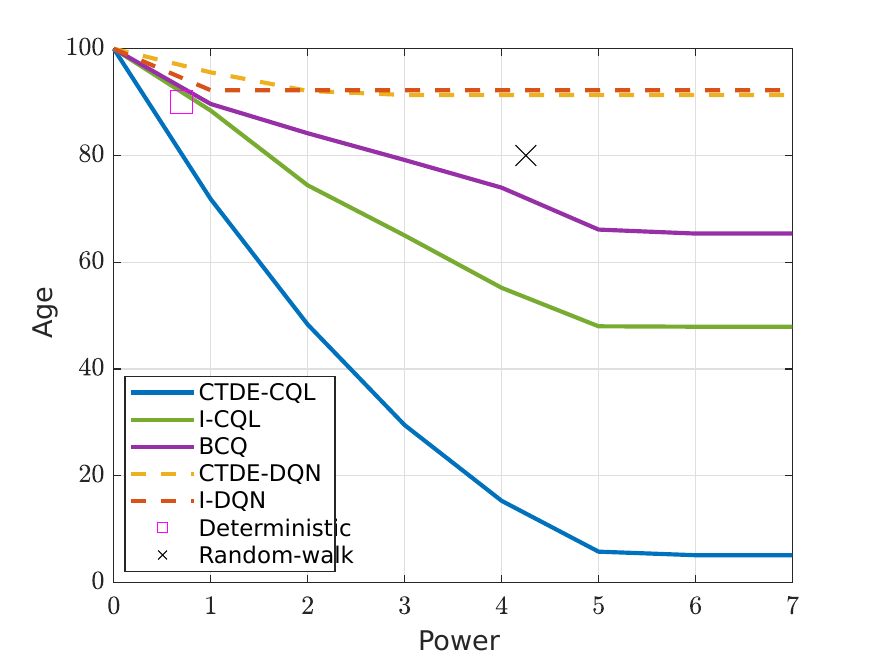} \vspace{2mm}
    \caption{The achieved Age and transmission power using different power factors after offline training.}
    \vspace{0mm}
    \label{UAV_Fig}
\end{figure}

\subsection{Case Study 2: UAV Trajectory Planning} \label{sec:UAV_usecase}
In this use case, we simulate a $1100 \: \text{m} \times 1100 \: \text{m}$ area with $15$ randomly deployed low-power IoT devices that need to transmit their data to $3$ UAVs flying at a fixed altitude of $100 \: \text{m}$. In all episodes, device positions and UAV initial positions are randomized, which reflects mobility and scalability in practical scenarios. The primary objective is to co-design the UAV trajectories to jointly minimize the age-of-information (AoI) and the transmission power of the devices~\cite{10753476}. Each UAV observes its current position and the AoI of each device (local observations) and selects an action comprising its movement direction $\{$north, south, east, west, hover$\}$ and the specific device to receive data from. The reward function is formulated as a combination of the average AoI across all devices and the total transmission power, scaled by a power factor to balance the trade-off between AoI and transmission power. We generate an offline dataset using $20 \%$ of the experience obtained from online DQN agents. 

This experiment evaluates the achieved AoI at specific power levels for the proposed I-CQL and CTDE-CQL algorithms after training and during testing over $100$ time steps. We compare their performance against their counterparts, I-DQN and CTDE-DQN, which are built on a DQN architecture, as well as the deterministic model, where UAVs follow pre-determined paths connecting the devices, and the random-walk model, which selects actions randomly. In Fig.4, the axes represent the joint optimization outcome of average AoI and total transmission power, where the power factor in the reward function controls the trade-off. As shown in Fig.~\ref{UAV_Fig}, the deterministic model conserves energy by moving along fixed paths toward the devices but incurs very high AoI due to inefficiencies in timing. Similarly, the random-walk model incurs high AoI and power consumption by arbitrarily selecting actions at random. The baseline models, I-DQN and CTDE-DQN, struggle with the distributional shift in the offline datasets, leading to worse combined AoI and power outcomes than the deterministic and random-walk models. Similarly, BCQ results in a suboptimal age-power region. In contrast, the proposed I-CQL and CTDE-CQL algorithms effectively address the distributional shift issue, achieving low AoI and power jointly. Consistent with the previous experiment, CTDE-CQL outperforms I-CQL by leveraging Q-function sharing among agents.

\begin{figure}[t!]
    \centering    \includegraphics[width=0.8\columnwidth,trim={0cm 0 0cm 0},clip]{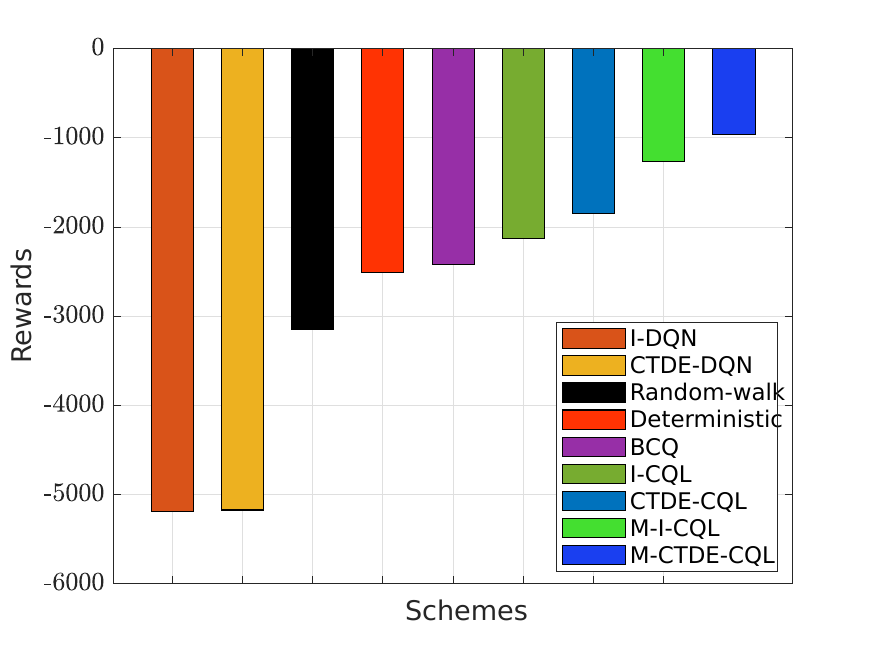} \vspace{2mm}
    \caption{The converged average rewards over $100$ unique test episodes on a new unseen task using a unique power factor.}
    \vspace{0mm}
    \label{Dynamic_UAV_Fig}
\end{figure}

\subsection{Case Study 3: UAV Trajectory Planning in Dynamic Networks} \label{sec:meta_UAV_usecase}
In this experiment, we extend the previous system model by defining multiple tasks (environments), where each task features a distinct objective determined by varying the power factor in the reward function, producing distinct AoI–power trade-offs~\cite{eldeeb2025multiagentmetaofflinereinforcementlearning}. This creates related but non-identical environments for meta-training, enabling the framework to capture transferable structures. The goal is to identify initial Q-function weights for each agent that enable rapid adaptation to new tasks with minimal training iterations and small datasets. The baseline models are similar to those in the previous experiment: I-DQN, CTDE-DQN, deterministic, and random-walk. Additionally, we include two meta-learning-enhanced models, M-I-CQL and M-CTDE-CQL, which integrate MAML with CQL. The evaluation measures the rewards achieved after $20$ training epochs using an offline dataset derived from only $5 \%$ of the experience collected by online DQN agents. In the implementation, each task is defined by a reward with different power factors, representing AoI and power trade-offs. In meta-training, agents adapt their Q-networks to sampled tasks, and the updated parameters are aggregated to refine the initialization. In meta-testing, a new task is introduced, and agents fine-tune from this initialization. This setup shows that MAML supports faster and more stable adaptation in MARL than random initialization.

As shown in Fig.~\ref{Dynamic_UAV_Fig}, I-DQN and CTDE-DQN fail to perform effectively due to the distributional shift problem, while deterministic and random-walk models also achieve relatively poor rewards. Similar to the previous experiments, BCQ achieves moderate suboptimal performance. The proposed I-CQL and CTDE-CQL algorithms demonstrate moderate convergence, constrained by the limited training epochs and small datasets. These models initialize the Q-function weights randomly. By contrast, M-I-CQL and M-CTDE-CQL achieve higher rewards than all other schemes by leveraging Q-network initialization across similar tasks (simulated here with $5$ training tasks). This initialization optimizes the weights for convergence within a few training steps. Among these, M-CTDE-CQL attains the best results, reaching optimal policies and outperforming M-I-CQL due to sharing Q-networks among agents.

\vspace{-1mm}

\section{Open Challenges} \label{sec:Challenges}
%\MS{400 words}

Despite the promising potential of offline MARL in the wireless domain, several challenges remain that need further exploration:
\begin{itemize}
\item \textbf{AI explainability:} As AI is deployed in real-time applications like autonomous drones, enhancing AI explainability is crucial. The challenge lies in developing transparent models that can explain decisions made by RL agents in dynamic environments without sacrificing performance. Future work should focus on interpretable models and real-time explanations that maintain clarity while adapting to changes in the environment.

\item \textbf{Scalability of offline data:} As the number of agents and network entities grows, ensuring the collection of diverse and representative offline data remains a significant challenge. The scalability and integration of data collection mechanisms into the training process must be addressed. In addition, sensitivity analysis of dataset size and task sampling strategies should be included, as both may introduce bias or affect stability in practical implementations.

\item \textbf{Hybrid offline-online learning:} While fully offline MARL ensures safe training, real-world 6G environments often require online fine-tuning to adapt to unseen conditions. Combining offline pre-training with lightweight online adaptation is especially relevant for edge computing and UAV scenarios, where dynamics such as mobility, interference, and workload can change rapidly. Future research should explore mechanisms such as continuous dataset refreshing and hybrid offline–online learning to ensure policy relevance after deployment.

\item \textbf{Task distribution:} In meta-learning, it is often assumed that tasks are sampled from a similar distribution, which may not always hold in real-world applications. The challenge lies in developing meta-learning models that can generalize effectively when tasks vary significantly, especially in dynamic environments. Future research should focus on relaxing this assumption and creating methods that allow meta-learning algorithms to adapt to diverse and evolving task distributions without losing performance.

\item \textbf{Safety and exploration:} Although offline MARL algorithms like CQL address some safety concerns by limiting the influence of unseen experiences, further research is needed to ensure these algorithms remain safe as they evolve and adapt to dynamic environments.

\item \textbf{Generative AI in offline data collection:} Generative AI can create synthetic data for offline reinforcement learning, but ensuring the quality and diversity of this data is a significant challenge~\cite{chen2024deepgenerativemodelsoffline}. Future research should focus on improving generative models to produce realistic, diverse datasets that capture rare events and avoid biases, especially for complex, multi-agent environments.

\item \textbf{Foundation model–enhanced RL:} Integrating large pre-trained models, such as LLMs and multi-modal foundation models, with RL can enhance agents’ ability to interpret semantic information and intent-aware instructions in 6G networks. This integration is particularly relevant for mission-critical, low-latency scenarios, but scaling to multi-agent systems remains a challenge. Future research should explore communication protocols and efficient adaptation without extensive retraining.
\end{itemize}

\vspace{-1mm}

\section{End Line} \label{sec:discussion}
%\MS{100 words}
This article highlights the transformative potential of offline MARL in addressing the challenges posed by next-generation wireless networks, including beyond 5G and 6G. By leveraging a novel offline MARL framework built on CQL and further augmented with meta-learning techniques, we present a scalable and adaptable solution for dynamic, multi-agent environments. This approach eliminates the need for costly and unsafe online data collection while enabling fast policy optimization across diverse scenarios.

Through comprehensive case studies, including radio resource management and UAV trajectory planning in static and dynamic network settings, we demonstrate the efficacy of the proposed framework. The results underscore its ability to effectively optimize complex, high-dimensional problems, outperforming traditional benchmarks and existing MARL approaches. In particular, the integration of meta-learning showcases promising advancements in enabling fast adaptation to new tasks using limited training data and computational resources.

Despite these advancements, several open challenges remain. Offline MARL still faces issues related to data scalability, ensuring robust generalization across diverse network conditions, and balancing safety with exploration. The limitations of current offline datasets and the inherent complexities of multi-agent systems highlight the need for innovative solutions to enhance algorithmic efficiency and performance further. Additionally, the reliance on assumptions about task distribution and the integration of explainability in decision-making processes are critical areas for future exploration. Future work may incorporate more challenging network aspects, such as time-varying inter-agent communication links, to further validate the proposed framework.

In conclusion, offline MARL represents a significant leap forward in applying artificial intelligence to wireless networks, providing a pathway to meet the demands of increasingly sophisticated and resource-intensive use cases. By continuing to address the outstanding challenges, offline MARL can play a pivotal role in shaping the future of wireless technologies and driving innovation across domains such as smart cities, autonomous systems, and the industrial Internet of Things.

\bibliographystyle{IEEEtran}
\bibliography{ref}

@ARTICLE{10753476,
  author={Eldeeb, Eslam and Sifaou, Houssem and Simeone, Osvaldo and Shehab, Mohammad and Alves, Hirley},
  journal={IEEE Transactions on Cognitive Communications and Networking}, 
  title={Conservative and Risk-Aware Offline Multi-Agent Reinforcement Learning}, 
  year={2024},
  volume={},
  number={},
  pages={1-1},
  doi={10.1109/TCCN.2024.3499357}}

@misc{levine2020offl,
      title={Offline Reinforcement Learning: Tutorial, Review, and Perspectives on Open Problems}, 
      author={Sergey Levine and Aviral Kumar and George Tucker and Justin Fu},
      year={2020},
      eprint={2005.01643},
      archivePrefix={arXiv},
      primaryClass={cs.LG},
      url={https://arxiv.org/abs/2005.01643}, 
}

@inproceedings{kumar2020conservative,
 author = {Kumar, Aviral and Zhou, Aurick and Tucker, George and Levine, Sergey},
 booktitle = {Advances in Neural Information Processing Systems},
 pages = {1179--1191},
 publisher = {Curran Associates, Inc.},
 title = {Conservative {Q}-Learning for Offline Reinforcement Learning},
 url = {https://proceedings.neurips.cc/paper_files/paper/2020/file/0d2b2061826a5df3221116a5085a6052-Paper.pdf},
 volume = {33},
 year = {2020}
}

@ARTICLE{9144301,
  author={Chowdhury, Mostafa Zaman and Shahjalal, Md. and Ahmed, Shakil and Jang, Yeong Min},
  journal={IEEE Open Journal of the Communications Society}, 
  title={{6G} Wireless Communication Systems: Applications, Requirements, Technologies, Challenges, and Research Directions}, 
  year={2020},
  volume={1},
  number={},
  pages={957-975},
  doi={10.1109/OJCOMS.2020.3010270}}

@ARTICLE{10155733,
  author={Cheng, Peng and Chen, Youjia and Ding, Ming and Chen, Zhuo and Liu, Sige and Chen, Yi-Ping Phoebe},
  journal={IEEE Communications Magazine}, 
  title={Deep Reinforcement Learning for Online Resource Allocation in {IoT} Networks: Technology, Development, and Future Challenges}, 
  year={2023},
  volume={61},
  number={6},
  pages={111-117},
  doi={10.1109/MCOM.001.2200526}}

@ARTICLE{FINN,  
    author={Chelsea Finn, Pieter Abbeel,Sergey Levine},  
    journal={34th International Conference on Machine Learning},
    title={Model-agnostic meta-learning for fast
    adaptation of deep networks}, 
    year={2017},  
    volume={70},  
    pages={1126-1135}
}

@misc{chen2024deepgenerativemodelsoffline,
      title={Deep Generative Models for Offline Policy Learning: Tutorial, Survey, and Perspectives on Future Directions}, 
      author={Jiayu Chen and Bhargav Ganguly and Yang Xu and Yongsheng Mei and Tian Lan and Vaneet Aggarwal},
      year={2024},
      eprint={2402.13777},
      archivePrefix={arXiv},
      primaryClass={cs.LG},
      url={https://arxiv.org/abs/2402.13777}, 
}

@ARTICLE{9372298,
  author={Feriani, Amal and Hossain, Ekram},
  journal={IEEE Communications Surveys \& Tutorials}, 
  title={Single and Multi-Agent Deep Reinforcement Learning for AI-Enabled Wireless Networks: A Tutorial}, 
  year={2021},
  volume={23},
  number={2},
  pages={1226-1252},
  doi={10.1109/COMST.2021.3063822}}

@ARTICLE{9738819,
  author={Li, Tianxu and Zhu, Kun and Luong, Nguyen Cong and Niyato, Dusit and Wu, Qihui and Zhang, Yang and Chen, Bing},
  journal={IEEE Communications Surveys \& Tutorials}, 
  title={Applications of Multi-Agent Reinforcement Learning in Future Internet: A Comprehensive Survey}, 
  year={2022},
  volume={24},
  number={2},
  pages={1240-1279},
  doi={10.1109/COMST.2022.3160697}}

@ARTICLE{9084325,
  author={Lee, Donghwan and He, Niao and Kamalaruban, Parameswaran and Cevher, Volkan},
  journal={IEEE Signal Processing Magazine}, 
  title={Optimization for Reinforcement Learning: From a single agent to cooperative agents}, 
  year={2020},
  volume={37},
  number={3},
  pages={123-135},
  doi={10.1109/MSP.2020.2976000}}

@ARTICLE{8807386,
  author={Cui, Jingjing and Liu, Yuanwei and Nallanathan, Arumugam},
  journal={IEEE Transactions on Wireless Communications}, 
  title={Multi-Agent Reinforcement Learning-Based Resource Allocation for UAV Networks}, 
  year={2020},
  volume={19},
  number={2},
  pages={729-743},
  doi={10.1109/TWC.2019.2935201}}

@inproceedings{NEURIPS2022_01d78b29,
 author = {Tseng, Wei-Cheng and Wang, Tsun-Hsuan Johnson and Lin, Yen-Chen and Isola, Phillip},
 booktitle = {Advances in Neural Information Processing Systems},
 pages = {226--237},
 publisher = {Curran Associates, Inc.},
 title = {Offline Multi-Agent Reinforcement Learning with Knowledge Distillation},
 volume = {35},
 year = {2022}
}

@misc{eldeeb2025offlinemultiagentreinforcementlearning,
      title={An Offline Multi-Agent Reinforcement Learning Framework for Radio Resource Management}, 
      author={Eslam Eldeeb and Hirley Alves},
      year={2025},
      eprint={2501.12991},
      archivePrefix={arXiv},
      primaryClass={cs.MA},
      url={https://arxiv.org/abs/2501.12991}, 
}

@misc{eldeeb2025multiagentmetaofflinereinforcementlearning,
      title={Multi-Agent Meta-Offline Reinforcement Learning for Timely {UAV} Path Planning and Data Collection}, 
      author={Eslam Eldeeb and Hirley Alves},
      year={2025},
      eprint={2501.16098},
      archivePrefix={arXiv},
      primaryClass={cs.MA},
      url={https://arxiv.org/abs/2501.16098}, 
}

@ARTICLE{10173668,
  author={Gan, Deqiao and Ge, Xiaohu and Li, Qiang},
  journal={IEEE Internet of Things Journal}, 
  title={An Optimal Transport-Based Federated Reinforcement Learning Approach for Resource Allocation in Cloud–Edge Collaborative {IoT}}, 
  year={2024},
  volume={11},
  number={2},
  pages={2407-2419},
  doi={10.1109/JIOT.2023.3292368}}

\vspace{0mm}

\section*{Biographies}
\footnotesize

\vspace{0mm}

\noindent\textbf{ESLAM ELDEEB} received a B.Sc. degree in electrical engineering from Alexandria University, Egypt, in 2019 and an M.Sc. degree from the University of Oulu, Finland, in 2021, where he obtained his doctoral degree in 2022. Eslam was a visiting researcher at King's College London during 2023. He is actively working on massive connectivity and ultrareliable low-latency communication. His research interests are machine-type communication and machine learning for wireless communication networks. \\

\vspace{-1mm}

\noindent \textbf{HIRLEY ALVES} (S’11–M’15) is an Associate Professor on Machine-type Wireless Communications at the Centre for Wireless Communications, University of Oulu, where he leads the Massive Wireless Automation Theme in the 6G Flagship Program. He earned his B.Sc. and M.Sc. in electrical engineering from the Federal University of Technology-Paraná, Brazil, and holds a dual D.Sc. from the University of Oulu and UTFPR. His research interests are massive and critical MTC, satellite IoT, distributed processing and learning, beyond 5G and 6G technologies.  

% PHOTO and bio here: https://unioulu.sharepoint.com/:w:/r/sites/MTCgroup/Shared%20Documents/Templates/hirley-bios-photos.docx?d=w6463213c30974fe291ad53854aaf5296&csf=1&web=1&e=BcJH8c 

\end{document}